\documentclass[journal]{IEEEtran}
\usepackage[numbers,sort&compress]{natbib}
\usepackage{amsmath,amsfonts}
\usepackage{array}
\usepackage[hyphens]{url}
\usepackage{graphicx}
\usepackage{amsthm}
\usepackage{amssymb}
\usepackage{multirow}
\usepackage{makecell}
\usepackage{xcolor}
\usepackage{booktabs}
\usepackage{threeparttable}
\usepackage{breakurl}
\usepackage{bm}
\usepackage{color}
\usepackage{mathrsfs}
\usepackage{float}
\begin{document}

\title{HDA-SELD: Hierarchical Cross-Modal Distillation with Multi-Level Data Augmentation for Low-Resource Audio-Visual Sound Event Localization and Detection}

\author{Qing Wang,~\IEEEmembership{Member,~IEEE}, Ya Jiang, Hang Chen,~\IEEEmembership{Member,~IEEE}, Sabato Marco Siniscalchi,~\IEEEmembership{Senior Member,~IEEE}, Jun Du,~\IEEEmembership{Senior Member,~IEEE}, and Jianqing Gao
\thanks{Q. Wang, Y. Jiang, H. Chen, and J. Du are with University of Science and Technology of China, Hefei 230027, China (e-mail: qingwang2@ustc.edu.cn, yajiang@mail.ustc.edu.cn, hangchen@ustc.edu.cn,  jundu@ustc.edu.cn).}
\thanks{S. M. Siniscalchi is with University of Palermo, Palermo 90133, Italy (e-mail: sabatomarco.siniscalchi@unipa.it).}
\thanks{J. Gao is with iFLYTEK Research, Hefei 230088, China (email: jqgao@iflytek.com).}}%



\maketitle

\begin{abstract}
This work presents HDA-SELD, a unified framework that combines hierarchical cross-modal distillation (HCMD) and multi-level data augmentation to address low-resource audio-visual (AV) sound event localization and detection (SELD). An audio-only SELD model acts as the teacher, transferring knowledge to an AV student model through both output responses and intermediate feature representations. To enhance learning, data augmentation is applied by mixing features randomly selected from multiple network layers and associated loss functions tailored to the SELD task. Extensive experiments on the DCASE 2023 and 2024 Challenge SELD datasets show that the proposed method significantly improves AV SELD performance, yielding relative gains of 21\%–38\% in the overall metric over the baselines. Notably, our proposed HDA-SELD achieves results comparable to or better than teacher models trained on much larger datasets, surpassing state-of-the-art methods on both DCASE 2023 and 2024 Challenge SELD tasks.
\end{abstract}

\begin{IEEEkeywords}
Sound event localization and detection, audio-visual fusion, data augmentation, knowledge distillation
\end{IEEEkeywords}

\section{Introduction}
Sound event localization and detection (SELD) \cite{butko2011two,chakraborty2014sound,hirvonen2015classification,adavanne2018sound} integrates two fundamental auditory tasks: identifying the temporal occurrence of specific acoustic events and estimating their spatial locations in three-dimensional (3D) space. As a pivotal component of computational auditory scene analysis, SELD has significant applications in audio surveillance, autonomous vehicles, bioacoustic monitoring, and so on.

Recently, deep neural networks (DNNs) have shown promising performance in SELD compared to traditional signal processing-based approaches. Early works \cite{butko2011two,chakraborty2014sound} employed Gaussian mixture model (GMM) - hidden Markov model (HMM) classifiers for sound event recognition and utilized the steered response power phase transform (SRP-PHAT) approach or steered beamforming sound model-based (SBSMB) technique for source localization. Hirvonen \cite{hirvonen2015classification} proposed the first DNN-based method for SELD, demonstrating that the application of a convolutional neural network (CNN) is more suitable for adapting to specific situations than hand-engineering and optimization. Adavanne et al. \cite{adavanne2018sound} introduced a convolutional recurrent neural network (CRNN) to address the SELD task, featuring two parallel branches that perform sound event detection (SED) and direction-of-arrival (DOA) estimation simultaneously under diverse acoustic scenarios. Subsequently, researchers began to investigate the impact of acoustic features \cite{cao2019polyphonic,politis2020overview,nguyen2022salsa,jacome2023sound}, output representation formats \cite{cao2021improved,shimada2022multi,kim2023ad}, network architectures \cite{hu2022track,wang2023four,shul2024cst,mu2024mff,wang2025fa3}, and data augmentation methods \cite{mazzon2019first,he2021neural,wu2023haac,zhang2024automated} on SELD. Several researchers explored the use of pre-trained model to improve SELD performance \cite{santos2024w2v,he2025adapting,nozaki2025source,hu2025pseldnets}.

Auditory and visual modalities play a crucial role in human communication and scene understanding. Audio-visual (AV) learning, which mimics human perceptual capabilities, has flourished in both academia and industry, facilitating advancements in speech recognition \cite{song2022multimodal}, speech enhancement \cite{chen2021correlating}, action recognition \cite{zhang2022audio}, question answering \cite{yun2021pano}, and other areas. By leveraging complementary information from both modalities, audio-visual learning overcomes the limitations inherent in unimodal approaches. Notably, visual representations of sound events remain unaffected by environmental noise and room reverberation, making visual cues particularly valuable for enhancing SELD performance in complex acoustic environments. The emerging field of audio-visual fusion for SELD has attracted increasing research attention. Both the Detection and Classification of Acoustic Scenes and Events (DCASE) Challenge and the International Conference on Acoustics, Speech and Signal Processing (ICASSP) introduced the first AV SELD tasks \cite{shimada2023starss23,gramaccioni2024l3das23} in 2023, with the aim of leveraging visual cues for enhancing model performance. Compared to unimodal audio datasets, audio-visual data experiences severe scarcity issues, presenting unique challenges for the development of robust multi-modal systems. 

When addressing the challenging AV SELD tasks that suffer from severe data scarcity, carefully designed data augmentation methods may serve as a fundamental approach to enhance the model's generalization capability. However, few studies to date have specifically addressed online audio-visual data augmentation for SELD. In our previous work \cite{jiang2024exploring}, we demonstrated the effectiveness of the classical Mixup method \cite{zhang2018mixup} for both audio-only and audio-visual SELD tasks. While most existing data augmentation methods \cite{zhang2018mixup,verma2019manifold,yun2019cutmix,kim2020puzzle,faramarzi2022patchup} were designed for image classification tasks, their direct application to regression tasks, such as DOA and distance estimation, may introduce spatial label misalignment issues. 

On the other hand, knowledge distillation (KD) was independently proposed by Huang et al. \cite{huang2013cross} and Hinton et al. \cite{hinton2015distilling}, and has since emerged as a widely adopted technique across a broad range of domains, including model compression \cite{kim2016sequence}, object detection \cite{zhang2020improve}, and self-supervised learning \cite{song2023multi}. Various categories of knowledge have been explored for distillation, including logits \cite{zhao2022decoupled}, feature maps \cite{yim2017gift}, and sample relationships \cite{yang2022cross}. Several studies have focused on cross-modal knowledge distillation (CMKD) \cite{li2019improving,liu2023emotionkd,gao2024cross,fan2025cross}, which facilitates the transfer of knowledge between different modalities. The primary objective of CMKD is to leverage the implicit cross-modal relational knowledge from a teacher model to enhance the performance of a student model, thereby addressing challenges related to limited labeled data and modality misalignment in target domains. However, in this work, an audio-only SELD model trained on a large dataset acts as the teacher, transferring knowledge to an AV student model trained on a limited dataset. 

Considering the challenges associated with obtaining high-quality audio-visual data in real-world scenarios, we propose a novel hierarchical cross-modal distillation (HCMD) framework that simultaneously transfers informative patterns from both output responses and hidden layer representations to address low-resource AV SELD in this study. In comparison to audio-visual data collection, acquiring or simulating realistic audio-only data is relatively more feasible, enabling us to train a robust audio-only SELD model to serve as the teacher. The rich knowledge embedded in the audio model, which is trained on extensive audio data, is then transferred to an audio-visual student model that is trained on limited multi-modal data. Furthermore, we conduct a comprehensive investigation of multi-level data augmentation specifically for AV SELD tasks. To enhance the diversity of training samples, multi-granularity features from multiple network layers are dynamically mixed to synthesize new training data. Crucially, instead of mixing localization labels directly, we propose mixing the corresponding loss functions to better preserve spatial relationships for the source localization task. The proposed multi-level data augmentation is systematically applied across three components: (1) audio-only SELD training, (2) audio-visual SELD training, and (3) the HCMD process. 

The main contributions of this work are as follows: 
\begin{itemize}
    \item We investigate the effectiveness of hierarchical cross-modal distillation for AV SELD. Rich information from the output responses and intermediate features of the audio model is utilized to supervise the learning of the audio-visual model, effectively compensating for data scarcity in multi-modal training scenarios.
    \item We propose multi-level data augmentation techniques tailored for AV SELD, which involve randomly mixing point-level or patch-level features from multiple network layers, while preserving spatial consistency through loss interpolation rather than direct label manipulation.
    \item We conduct comprehensive experiments on the DCASE 2023 and 2024 Challenge SELD datasets, demonstrating the effectiveness of the HDA-SELD framework, which combines hierarchical cross-modal distillation with multi-level data augmentation. Furthermore, the proposed HDA-SELD achieves performance that is comparable to or superior to that of the teacher model, outperforming our previous first-place approaches. 
\end{itemize}

\section{Proposed Method}

\begin{figure*}[t]
\centering
\includegraphics[width=1.81\columnwidth]{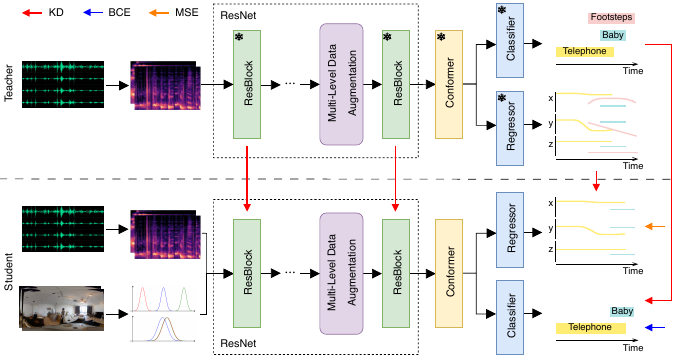}
\caption{Hierarchical cross-modal distillation framework. During training, the audio-only model (teacher) transfers its rich acoustic knowledge to the audio-visual model (student) through hierarchical cross-modal distillation. At reference, the student model leverages visual inputs to perform SELD.}
\label{fig1}
\end{figure*}

In this section, we present a detailed description of our HCMD method for AV SELD. Specifically, Section \ref{overall} introduces the overall architecture of the proposed framework. Section \ref{hcmd} elaborates on the core components of hierarchical cross-modal distillation. Section \ref{mlda} details the multi-level data augmentation strategy, which dynamically mixes point-level or patch-level representations across multiple encoder layers. Section \ref{loss} outlines the training objectives.

\subsection{Overall Architecture}
\label{overall}

Owing to the challenges associated with data collection and annotation costs, existing audio-visual datasets for SELD tasks are significantly smaller in scale compared to audio-only datasets \cite{brousmiche2020secl,nagatomo2022wearable,guizzo2022l3das22,shimada2023starss23,gramaccioni2024l3das23}. To address this limitation, our study aims to enhance the performance of audio-visual models through hierarchical cross-modal transfer from the audio modality to the audio-visual modality. As illustrated in Fig.~\ref{fig1}, we propose utilizing an audio-only SELD model as the teacher model and an audio-visual SELD model as the student model. The proposed method employs a two-stage training strategy, consisting of a pretraining phase for the teacher model, and a cross-modal knowledge transfer phase for the student model. In the first stage, we train the teacher model using large-scale audio-only data to capture robust acoustic representations. In the second stage, the weights of the teacher model are frozen, and its learned generalized knowledge serves as a regularization constraint to guide the audio-visual student model. Furthermore, we introduce a multi-level data augmentation strategy that dynamically mixes representations from different network layers, either at the point-level or the patch-level. This approach is specifically designed to improve the performance of both the teacher and student models.

\subsection{Hierarchical Cross-modal Distillation}
\label{hcmd}

The proposed method aims to develop a single-modality teacher model to enhance the performance of a multi-modal student model through knowledge distillation. As illustrated in Fig.~\ref{fig1}, the proposed approach comprises three key components: an audio model (SELD-Teacher), an audio-visual model (SELD-Student), and the knowledge distillation method. In this HCMD framework, the teacher model is trained on large-scale audio data, enabling it to capture fine-grained acoustic features and robust contextual relationships. Despite being constrained by limited audio-visual training data, the student model effectively inherits robust acoustic representation abilities through knowledge distillation while simultaneously establishing cross-modal correlations from visual inputs. In this subsection, we will elaborate on these three key components.

\subsubsection{SELD-Teacher}
In contrast to previous knowledge distillation paradigms, where multi-modal models guide single-modality models \cite{liu2023emotionkd,fan2025cross}, we reverse the knowledge flow by training a teacher model exclusively on a large-scale audio dataset. As shown in Fig.~\ref{fig1}, SELD-Teacher \cite{wang2023four} comprises a residual network (ResNet) module, a Conformer module, a classifier for SED, and a regressor for sound source localization (SSL). The model takes log-Mel spectrogram and intensity vector features as input, utilizing an 18-layer ResNet architecture as the audio encoder, with an 8-layer Conformer module to model both long-term and short-term contextual relationships. Subsequently, the temporal features are fed into two parallel branches to perform joint sound event localization in both temporal and spatial domains via multi-task learning. 

The audio data is provided in the first-order Ambisonics (FOA) format, which includes four channels of signals: w, x, y, and z. These four channel signals encode omni-directional, $x$-directional, $y$-directional, and $z$-directional spatial sound information. The active intensity vector quantifies the flow of acoustic energy in 3D space, which is used for DOA estimation \cite{perotin2019crnn}. The acoustic features used in this study are derived by concatenating log-Mel spectrograms and intensity vectors \cite{wang2023four}.

\begin{figure*}
\centering
\includegraphics[width=1.9\columnwidth]{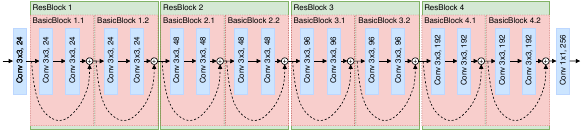}
\caption{Illustration of ResNet. The convolutional layer parameters are denoted as ``Conv $\left<\mathrm{kernel\ size}\right>$, $\left<\mathrm{number\ of\ channels}\right>$".}
\label{fig2}
\end{figure*}

The audio encoder utilizes an 18-layer ResNet to extract local shift-invariant features, as depicted in Fig.~\ref{fig2}. ResNet consists of four residual blocks (ResBlocks), each comprising two basic blocks (BasicBlocks), where every basic block includes two convolutional layers (Convs). Each Conv layer employs 2D kernels of size 3 × 3, along with batch normalization and a rectified linear unit (ReLU) activation function. Max-pooling is applied after the first three ResBlocks. 

\begin{figure}
\centering
\includegraphics[width=0.68\columnwidth]{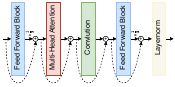}
\caption{Illustration of one Conformer layer.}
\label{fig3}
\vspace{-0.1cm}
\end{figure}

In order to model both local and global context dependencies in audio sequences, we employ an 8-leyer Conformer architecture, which combines convolution and transformer layers \cite{gulati2020conformer}. The Conformer architecture has been applied to SELD tasks, achieving state-of-the-art performance \cite{wang2023four,jiang2024exploring,dong2025exp}. As shown in Fig.~\ref{fig3}, each Conformer layer comprises two feed forward blocks that sandwich a multi-head attention block and a convolution block. All four blocks begin with layer normalization operations. In the multi-head attention block, each frame of the input interacts with other frames to capture long-term dependencies. The convolution layers effectively extract local fine-grained features. In the feed forward blocks, the features of each frame are projected into a high-dimensional space, thereby enhancing the model's representational capacity. 

Finally, the context representation is temporally downsampled to match the label's time resolution. Two parallel branches are employed for the joint prediction of both sound event categories and their spatial locations, respectively. Both the classifier and the regressor are composed of two FC layers. 

In summary, the proposed SELD-Teacher, trained on large-scale audio datasets, demonstrates robust predictive capabilities by effectively extracting high-level features and contextual temporal representations. The derived high-quality embeddings serve as effective knowledge to guide multi-modal learning, significantly improving the model's generalization performance in data-scarce scenarios.

\subsubsection{SELD-Student}

The audio-visual SELD student model, equipped with limited paired visual inputs, shares an identical backbone architecture with the teacher model. Specifically, the backbone comprises a ResNet audio encoder, a Conformer context module, a classification head, and a regression head. The main difference from the teacher model is the multi-modal input. The baseline of the DCASE 2023 Challenge Task 3 \cite{shimada2023starss23} used a object detection module to outputs bounding boxes of potential objects on target classes, which were demonstrated to perform better than CNN-based visual features \cite{he2016deep}. Following the baseline, we utilize a keypoint detection model to extract visual features. The visual features are converted from mouth coordinates into Gaussian vectors, which are explicitly designed to model spatial location information. 

\begin{figure}
\centering
\includegraphics[width=0.95\columnwidth]{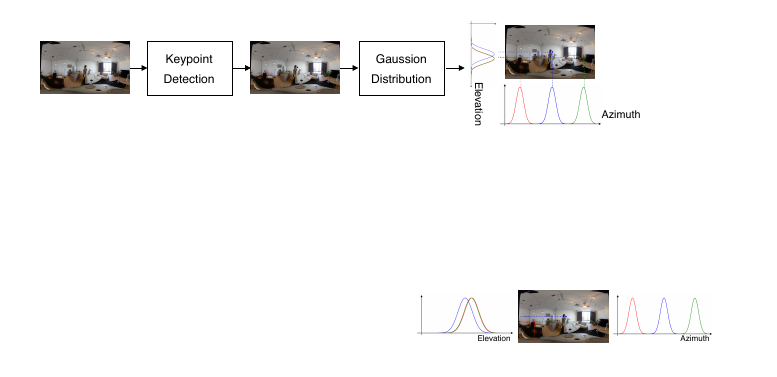}
\caption{Illustration of the visual encoded Gaussian vectors.}
\label{fig4}
\vspace{-0.1cm}
\end{figure}

The acoustic features are identical to those used in training the teacher model. For video data, we uniformly sample 10 frames per second to extract visual features. Using HRNet \cite{sun2019deep} as the keypoint detection model, we extract the mouth pixel coordinates $(u_s, v_s)$ for the $s$-th speaker in each frame. The mouth position is then encoded into two Gaussian-like vectors \cite{shimada2023starss23,qian2022audio}, representing probability likelihoods along the horizontal and vertical axes of the image, as depicted in Fig.~\ref{fig4}, characterizing the spatial presence of the $s$-th speaker,
\begin{equation}
    \rho^{s}_{\mathrm{azi}}(u) = \exp\left(\frac{-|u - u_s|^2}{\sigma_u^2}\right)
\end{equation}
\begin{equation}
    \rho^{s}_{\mathrm{ele}}(v) = \exp\left(\frac{-|v - v_s|^2}{\sigma_v^2}\right)
\end{equation}
where the means $u_s$ and $v_s$ denote the normalized pixel coordinates of the mouth keypoint along the horizontal and vertical axes for the $s$-th speaker, respectively. We consider a maximum of $S=6$ concurrent speakers in each video frame. If the detected number of speakers $S'$ ($S'$=3 in Fig.~\ref{fig4}) is less than six, we pad the visual feature with $2×(S-S')$ zero tensors to maintain consistent dimensionality. The horizontal and vertical vectors of all speakers are concatenated to form the visual feature input.

Subsequently, the visual feature is concatenated with the acoustic feature to form the multi-modal input. A similar backbone, as depicted in Fig.~\ref{fig1}, is applied to output the predicted sound probability and the Cartesian coordinate of SELD-Student. Crucially, the student model's weights are initialized with those of the pre-trained teacher, providing a powerful starting point instead of training from scratch. 

\subsubsection{Knowledge Distillation}
There exist some research on knowledge distillation frameworks \cite{li2019improving,liu2023emotionkd,fan2025cross} for multi-modal learning tasks. However, these KD frameworks \cite{liu2023emotionkd,fan2025cross} typically treat multi-models as the teacher and unimodal models as the student to address the potential issue of missing modalities in real scenarios, where the distillation process for intermediate features is driven by a simple mean squared error (MSE) loss. Li et al. \cite{li2019improving} proposed a cross-modal teacher-student learning framework to enhance audio-visual speech recognition (AVSR), in which the teacher model was trained on large-scale audio data and only output response distillation is used. Our approach differs from these CMKD frameworks in two key aspects. First, we address the data scarcity issue in multi-modal learning tasks by employing a model trained on large audio dataset as the teacher. Second, the distillation method designed to enhance the intermediate representations of the student model dynamically fuses information from multiple network stages, unlike the single-level distillation approaches commonly used in existing CMKD frameworks.

\begin{figure*}
  \centering
  \begin{minipage}[b]{0.8\linewidth}
  \centering
  \includegraphics[width=0.9\linewidth]{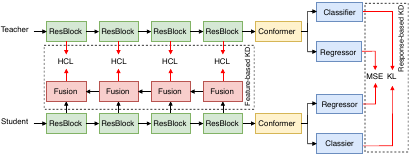}
  {\centerline{\scriptsize{(a)}}}
  \end{minipage}
  \label{fig:fig5}
  \hfill
  \begin{minipage}[b]{0.19\linewidth}
  \centering
  \includegraphics[width=0.85\linewidth]{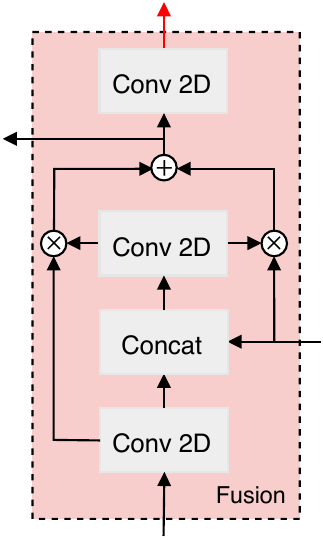}
  {\centerline{\scriptsize{(b)}}}
  \end{minipage}
  \label{fig:fig6}

  \centering
  \vspace{-0.4cm}
  \caption{(a) Detailed implementation of the proposed hierarchical cross-modal distillation, consisting response-based and feature-based KD methods. (b) The fusion method applied in the feature-based KD.}
  \label{fig:hcmd}
\end{figure*}

Specifically, the proposed HCMD method in this study aims to leverage a teacher model trained on large-scale audio data, transferring its distilled knowledge to the multi-modal student model. In our previous work \cite{jiang2024exploring}, we proposed transferring knowledge from the output responses of the teacher model to the student model. In this study, we extend the distillation approach to hierarchical cross-modal distillation, migrating rich information not only from the output responses but also from the intermediate features. 

As illustrated in Fig.~\ref{fig:hcmd}.(a), response-based KD (RKD) and feature-based KD (FKD) are implemented in the ResNet-Conformer architecture for SELD. An intuitive and straightforward way for knowledge distillation is using the response outputs of the audio teacher model to transfer effective knowledge to the response outputs of the audio-visual student model. We employ Kullback-Leibler (KL) as the distillation loss for the SED task and MSE for the SSL task. These two distillation losses guide the training of the student model, enabling it to emulate the teacher's behavior and consequently achieve improved performance. The RKD loss is expressed as,
\begin{align}
\mathcal{L}^{\text{rkd}} = &\beta_1 \times \mathcal{L}_{\text{sed}}^{\text{rkd}}  + \beta_2 \times \mathcal{L}_{\text{ssl}}^{\text{rkd}}\\
\mathcal{L}_{\text{sed}}^{\text{rkd}} \!= &-\!\!\frac{1}{LN}\!\sum_{l,n}\! \!\Big[ p_{l,n}^{\text {t}} \!\log \!\frac{p_{l,n}^{\text {t}}}{p_{l,n}^{\text {s}}}\! + \!(1\! -\! p_{l,n}^{\text {t}})\! \log\frac{(1 - p_{l,n}^{\text {t}})}{(1 - p_{l,n}^{\text {s}})} \Big]\! \\
\mathcal{L}_{\text{ssl}}^{\text{rkd}} = &\frac{1}{LN}\sum_{l,n} \left\| (\mathbf{y}_{l,n}^{\text {t}} - \mathbf{y}_{l,n}^{\text {s}})p_{l,n}^{\text {t}} \right\|^2
\end{align}
where $p_{l,n}^{\text {t}}$ and $p_{l,n}^{\text {s}}$ represent the sound activity probabilities predicted by the teacher model and student model, respectively, for the $l$-th frame and $n$-th event class. The vectors $\mathbf{y}_{l,n}^{\text {t}}$ and $\mathbf{y}_{l,n}^{\text {s}}$ represent the Cartesian coordinates of sound source locations predicted by the teacher model and student model, respectively, for the $n$-th sound event class in the $l$-th time frame. Each location vector $\mathbf{y}_{l,n} \in \mathbb{R}^3$ is defined as $\mathbf{y}_{l,n}=(x,y,z)$ in 3D space. The weights $\beta_1$ and $\beta_2$ are used to balance the SED and SSL losses.

Inspired by \cite{chen2021distilling}, we propose an FKD approach, which utilizes multiple layers of ResNet in the teacher to supervise one layer in the student. This knowledge distillation approach employs a fusion method, illustrated in Fig.~\ref{fig:hcmd}.(b), to dynamically aggregate feature maps from different stages of the network. Assuming the intermediate feature at the $j$-th stage of the student model is $\mathbf{F}_j^{\text {s}}$, and the residual output of the fusion module at the $(j+1)$-th stage is $\mathbf{R}_{j+1}^{\text {s}}$, the fusion process at the $j$-th stage is implemented as follows,
\begin{align}
\tilde{\mathbf{F}}_j^{\text {s}} &= \text{Conv2D}(\mathbf{F}_j^{\text {s}})\\
\hat{\mathbf{F}}_j^{\text {s}} &= \text{Concat}(\tilde{\mathbf{F}}_j^{\text {s}},\mathbf{R}_{j+1}^{\text {s}}) \\
\mathbf{Z}_j^{\text {s}} &= \text{Conv2D}(\hat{\mathbf{F}}_j^{\text {s}})\\
\mathbf{R}_j^{\text {s}} &= \mathbf{Z}_j^{\text {s}}[:,0] \odot \tilde{\mathbf{F}}_j^{\text {s}} + \mathbf{Z}_j^{\text {s}}[:,1] \odot \mathbf{R}_{j+1}^{\text {s}}\\
\bar{\mathbf{F}}_j^{\text {s}} &= \text{Conv2D}(\mathbf{R}_j^{\text {s}})
\end{align}
where $\text{Conv2D}$ denotes the 2D convolution layer, $\mathbf{Z}_j^{\text {s}}$ denotes the attention map used to fuse features from different stages, $\mathbf{R}_j^{\text {s}}$ represents the residual output, and $\bar{\mathbf{F}}_j^{\text {s}}$ is the fusion output for feature-based KD loss calculation. Notably, for intermediate features at the top stage of the student model, the attention computation is omitted due to the absence of residual input. Only intermediate features from the top stage are used to compute both the residual and the fused outputs.

Another key component of the feature-based KD method is hierarchical context loss (HCL) \cite{chen2021distilling}. HCL transfers knowledge between differently sized feature maps of the teacher and student models through spatial pyramid pooling (SPP). SPP captures context information at different levels through multi-scale pooling to generate fixed-length feature representations. Let $\mathbf{F}_j^{\text {t}}$ denote the intermediate feature of the teacher model corresponding to the fused output feature of the $j$-th stage student model. We extract their hierarchical knowledge and employ an MSE loss function to align the student model's intermediate features with those of the teacher at different layers. The HCL for the FKD method is written as,
\begin{align}
\mathcal{L}^{\text{fkd}} \!&= \!\sum_j^{J-1} \sum_m^M \alpha_m \left\| \mathbf{M}_{j,m}^{\text {t}}\!\!-\!\mathbf{M}_{j,m}^{\text {s}}\right\|^2 \!+ \!\sum_j^J \!\left\| \mathbf{F}_j^{\text {t}}\!-\!\bar{\mathbf{F}}_j^{\text {s}}\right\|^2
\end{align}
where $\mathbf{M}_{j,m}^{\text {t}}$ and $\mathbf{M}_{j,m}^{\text {s}}$ denote the feature maps of the teacher and student, respectively, for the $j$-th stage and the $m$-th adaptive average pooling of the SPP. The parameter $\alpha_m$ represents the loss weight assigned to the $m$-th pooled feature map. Residual learning is used in the FKD method. The high-level features of the student are combined with the low-level features to mimic those of the teacher. In this way, the high-level features of the student learns the residual of the low-level features between the teacher and the student. 

\subsection{Multi-level Data Augmentation}
\label{mlda}
Multi-level data augmentation aims to mix features at varying granularities, randomly selected from multiple layers of the network, with loss interpolation instead of using target labels. Let $\mathbf{X}$ and $\mathbf{X}'$ denote two training samples. Given a DNN $h$, let $\phi_d$ represent the mapping from the input data to the hidden representation at the $d$-th layer, where $d \in \{ 0,1,...,D\}$ and $D$ denotes the number of eligible layers for feature mixing. The network can be represented as $h(\mathbf{X})=h_d(\phi_d(\mathbf{X}))$, where $h_d$ is the mapping from the hidden representation at layer $d$ to the output \cite{verma2019manifold}. The goal of the multi-level data augmentation is to generate a new training sample by combining features from multiple layers. The general feature mixing operation at layer $d$ is defined as follows,
\begin{align}
\tilde{\mathbf{X}}_d= \mathbf{M} \odot \phi_d(\mathbf{X}) + (\mathbf{1}-\mathbf{M}) \odot \phi_d(\mathbf{X}')
\end{align}
where $\phi_d(\mathbf{X}) \in \mathbb{R}^{T_d \times F_d \times C_d}$ and $\phi_d(\mathbf{X}')\in \mathbb{R}^{T_d \times F_d \times C_d}$ denote two intermediate features at layer $d$, $\odot$ is element-wise multiplication, $T_a$, $F_d$ and $C_d$ represent the dimensions of time, frequency, and channel at layer $d$, respectively, and $\mathbf{M} \in \{0,1 \}^{F_d \times C_d}$ denotes the mixing coefficient matrix. 

We explored several feature mixing methods, including mixing at either the point level or patch level, as well as mixing at the input layer or hidden layers. Methods such as Mixup \cite{zhang2018mixup}, ManifoldMixup \cite{verma2019manifold}, and CutMix \cite{verma2019manifold} can be viewed as special cases of the general feature mixing method. 

When mixing is performed at the point level, the mixing coefficient matrix satisfies $\mathbf{M}=\lambda \mathbf{I}$, where $\lambda \sim \text{Beta}(\alpha, \alpha)$ follows the Beta distribution and $\mathbf{I}$ is an all-ones matrix. In this case, Mixup \cite{zhang2018mixup} can be seen as a special case of input-layer feature mixing, while ManifoldMixup \cite{verma2019manifold} represents a specialized form of hidden-layer mixing. 

For patch-level feature mixing, the mixing coefficient matrix becomes a binary mask that determines which regions of the two feature maps are mixed. A patch defined by the bounding box coordinates $[f1,f2,c1,c2]$ is selected, with the mask values set to 0 within this region. The computation of this patch follows a procedure similar to CutMix \cite{verma2019manifold}, where the key steps involve calculating the patch dimensions $(r_f,r_c)$ based on the mixing coefficient $\lambda$ and the feature map size $(F_d,C_d)$, then determining the patch boundaries using uniformly sampled center coordinates. The calculation process is expressed as follows,
\begin{align}
r_f =F_d\sqrt{1-\lambda},r_c &=C_d\sqrt{1-\lambda} \\
f_1 = \max(0, c_f - r_f/2), f_2 &= \min(F_d, c_f + r_f/2) \\
c_1 = \max(0, c_c - r_c/2), \,\,c_2 &= \min(C_d, c_c + r_c/2)
\end{align}
where $\lambda \sim \text{Beta}(\alpha, \alpha)$ is sampled from the Beta distribution, the center coordinates $c_f \sim \mathcal{U}(0, F_d)$ and $\quad c_c \sim \mathcal{U}(0, C_d)$ are sampled from two uniform distributions. The variables $r_f$ and $r_c$ denote the width and height of the patch, respectively. The patch coordinates $(f1,f2,c1,c2)$ are then computed by adjusting the center coordinates to ensure that they remain within valid feature map boundaries.

Most data augmentation methods, such as Mixup \cite{zhang2018mixup}, ManifoldMixup \cite{verma2019manifold}, and CutMix \cite{yun2019cutmix}, were designed for image classification tasks, combining data and labels to expand the diversity of training examples. However, for a regression task like sound source localization, the label mixing operation may lead to incorrect targets for the synthesized samples. To address this, inspired by recent work \cite{vu2024supervision}, we propose using mixed loss as the supervision signal $\mathbf{S}$ instead of mixing the labels, which is more suitable for optimizing SELD tasks by preserving directional accuracy. Given the mixing coefficient $\lambda \sim \text{Beta}(\alpha, \alpha)$, the general mixing formulation of the augmented supervision signal is written as follows,
\begin{equation}
\tilde{\mathbf{S}} = \lambda \mathbf{S} + (1-\lambda)\mathbf{S}'
\end{equation}

In Mixup-based augmentation methods \cite{zhang2018mixup,verma2019manifold,yun2019cutmix}, the ground truths of SELD labels serve as the supervision signals. Subsequently, the augmented target labels corresponding to the mixed features are used to train SELD models. On the contrary, our proposed augmentation methods treat the loss errors $\mathcal{L}$ as the supervision signals. Equation (26) can be expressed in the form of augmented loss,
\begin{equation}
\tilde{\mathcal{L}} = \lambda \mathcal{L} + (1-\lambda)\mathcal{L}'
\end{equation}

A comparison of several data augmentation methods on mixing granularity, mixing layer, and mixing supervision is presented in Table~\ref{tab: table1}, which are investigated in this study.

\begin{table}[t]
\caption{A comparison of several data augmentation methods.}
\centering
\label{tab: table1} 
\setlength{\tabcolsep}{5pt}
\begin{tabular}{ccccccc}
    \toprule
    \multirow{2}*{Method} & \multicolumn{2}{c}{\raisebox{0.3\height}{Granularity}} & \multicolumn{2}{c}{\raisebox{0.3\height}{Layer}} & \multicolumn{2}{c}{\raisebox{0.3\height}{Supervision}} \\
    \cline{2-7}
       ~  & \raisebox{-0.3\height} {Point} & \raisebox{-0.3\height}{Patch} & \raisebox{-0.3\height} {Input} & \raisebox{-0.3\height} {Hidden} &  \raisebox{-0.3\height} {Label} & \raisebox{-0.3\height} {Loss}\\
        \midrule
        Mixup \cite{zhang2018mixup} & \checkmark & - & \checkmark & - & \checkmark & - \\
        LossMix \cite{vu2024supervision} & \checkmark & - & \checkmark & - & - & \checkmark \\
        ManifoldMixup \cite{verma2019manifold} & \checkmark & - & - & \checkmark & \checkmark & -  \\
        PointMix & \checkmark &-  & - & \checkmark & - & \checkmark \\
        CutMix \cite{yun2019cutmix} & - & \checkmark & \checkmark & - & \checkmark & -  \\
        PatchMix & - & \checkmark & - & \checkmark & - & \checkmark \\
        \bottomrule
\end{tabular}
\end{table}

\subsection{Training Objectives}
\label{loss}
The SELD task loss, denoted as $\mathcal{L}^{\text{s}}$, consists of binary cross-entropy (BCE) for SED and MSE for SSL, defined as follows,
\begin{align}
\mathcal{L}^{\text{s}} &= \,\,\eta_1 \times \mathcal{L}_{\text{sed}}^{\text{s}}  + \eta_2 \times \mathcal{L}_{\text{ssl}}^{\text{s}}\\
\mathcal{L}_{\text{sed}}^{\text{s}} &= -\!\frac{1}{LN}\!\sum_{l,n} \!\Big[ {\hat {p}_{l,n}} \log {p_{l,n}} \!+\! (1 \!- {\hat {p}_{l,n}}) \log (1\! - \!{p_{l,n}}) \Big]\\
\mathcal{L}_{\text{ssl}}^{\text{s}} &= \frac{1}{LN}\sum_{l,n} \left\| (\mathbf{\hat y}_{l,n} - \mathbf{y}_{l,n}){\hat {p}_{l,n}} \right\|^2
\end{align}
where $\{p_{l,n}, {\hat {p}_{l,n}}\}$ and $\{\mathbf{y}_{l,n}, \mathbf{\hat y}_{l,n}\}$ represent the model output and the ground truth for SED and SSL, respectively, for the $n$-th sound event at the $l$-th frame. The coefficients $\eta_1$ and $\eta_2$ are used to balance the contributions of the SED and SSL subtasks to the overall loss during training. 

The task loss is employed to train both SELD-Teacher and SELD-Student models. Concurrently, the multi-modal student model is required to learn from the audio teacher model to enhance SELD performance. We utilize the response-based KD and feature-based KD losses, denoted as $\mathcal{L}^{\text{rkd}}$ and $\mathcal{L}^{\text{fkd}}$, to align the outputs of the student model with those of the teacher model. Additionally, we investigate the impact of a feature-based KD method across various training data sizes. The hierarchical cross-modal distillation is discussed in detail in Section~\ref{hcmd}.

Overall, our proposed hierarchical cross-modal distillation method leverages a multi-task learning framework to address the challenges of low-resource AV SELD. The objective function is defined as follows,
\begin{equation}
\mathcal{L} = \mathcal{L}^{\text{s}} + \gamma_1 \times \mathcal{L}^{\text{rkd}}  + \gamma_2 \times \mathcal{L}^{\text{fkd}}
\end{equation}
where $\mathcal{L}^{\text{s}}$ represents the task loss, and $\gamma_1$ and $\gamma_2$ are balance factors used to adjust the weights of the response-based KD loss $\mathcal{L}^{\text{rkd}}$ and the feature-based KD loss $\mathcal{L}^{\text{fkd}}$.

\section{Experiments}

\subsection{Datasets}
 In this study, we conduct experiments on the development set of the Sony-TAu Realistic Spatial Soundscapes 2023 (STARSS23) dataset \cite{shimada2023starss23}. To validate the effectiveness of our proposed method, we perform evaluations on two closely related yet distinct tasks based on the STARSS23 dataset: (1) SELD without distance estimation \cite{shimada2023starss23} (corresponding to DCASE 2023 Challenge Task 3), and (2) 3D SELD incorporating sound source distance estimation \cite{krause2024sound} (corresponding to DCASE 2024 Challenge Task 3). For DCASE 2023 Task 3, the learning target of the SSL branch is represented by the DOA vector of the sound event. For DCASE 2024 Task 3, we integrate angular and distance information into absolute 3D coordinates by multiplying the DOA vector with the source distance. For more details, please refer to \cite{dong2025exp}.

The STARSS23 dataset consists of synchronized multi-channel audio and video recordings that capture diverse acoustic environments across multiple room settings. These recordings are accompanied by precise temporal and spatial annotations for various sound events. The audio data features two distinct 4-channel spatial recording formats: FOA and tetrahedral microphone array (MIC), with our experiments utilizing the FOA format. Video data is captured using a 360-degree camera operating at 29.97 fps with a resolution of 1920×960 pixels. The development set spans approximately 7.4 hours, including a dedicated training subset (dev-set-train, 90 clips) and a testing subset (dev-set-test, 78 clips). The dataset includes 13 distinct sound event categories, with temporal annotations provided at a resolution of 100 msec. The 13 sound classes are as follows: female speech, male speech, clapping, telephone, laughter, domestic sounds, footsteps, door, music, musical instruments, faucet, bell, and knock. The top five most frequent sound classes are male speech, female speech, music, domestic sounds, and laughter, which collectively account for over 90\% of the entire dataset. We utilize the officially partitioned training and testing subsets of the development set to train and test our models.

To expand the training set, the 20-hour synthetic DCASE dataset \cite{politis2022starss22} is combined with STARSS23. Audio channel swapping (ACS) \cite{wang2023four} and video pixel swapping (VPS) \cite{jiang2024exploring} methods are further employed to increase the data size by eightfold. The data size for each data set combination is illustrated is Table~\ref{tab: table4}.

\subsection{Experimental Setup}
For audio processing, STFT with a Hamming window of length 1024 samples and a 50\% overlap is used to extract a linear spectrogram for each of the four channels. Subsequently, a 64-dimensional log Mel-spectrogram is extracted for FOA data. Additionally, acoustic intensity is calculated at each of the 64 Mel-bands for channels x, y, and z of the FOA signal, resulting in three channels of intensity vector features. Therefore, there are a total of seven feature maps for FOA signals. For audio data segmented into 10-second intervals, the acoustic feature has a shape of $500\times 64\times 7$.

\begin{table*}[t]
    \centering
    \caption{Performance comparison of various on-line data augmentation methods for SELD and AV SELD tasks evaluated on the development set of the DCASE 2023 Challenge Task 3.}
    \label{tab: table2}
    \setlength{\tabcolsep}{9pt}
    \begin{tabular}{cccccc|ccccc}
    \toprule
    \multirow{2}*{Method} & \multicolumn{5}{c}{\raisebox{0.3\height}{SELD-Teacher (Audio-Only)}} & \multicolumn{5}{c}{\raisebox{0.3\height}{SELD-Student (Audio-Visual)}} \\
    \cline{2-11}
    ~  & \raisebox{-0.3\height} {$\text{ER}\!\downarrow$} 
    & \raisebox{-0.3\height}{$\text{F}\!\uparrow$} 
    & \raisebox{-0.3\height} {$\text{LE}\!\downarrow$} 
    & \raisebox{-0.3\height} {$\text{LR}\!\uparrow$} 
    & \raisebox{-0.3\height} {$\text{Score}\!\downarrow$} 
    & \raisebox{-0.3\height} {$\text{ER}\!\downarrow$} 
    & \raisebox{-0.3\height}{$\text{F}\!\uparrow$} 
    & \raisebox{-0.3\height} {$\text{LE}\!\downarrow$} 
    & \raisebox{-0.3\height} {$\text{LR}\!\uparrow$} 
    & \raisebox{-0.3\height} {$\text{Score}\!\downarrow$}\\
    \midrule
    NoAug &0.44 &0.540 &14.34$^{\circ}$ &0.660 &0.330 &0.58 &0.334 &18.80$^{\circ}$ &0.525 &0.456 \\
    Mixup &0.42 &0.571 &14.30$^{\circ}$ &0.672 &0.314 &0.51 &0.467 &16.93$^{\circ}$ &0.629 &0.377 \\
    LossMix &0.41 &0.586 &13.98$^{\circ}$ &0.682 &0.305 &0.49 &0.474 &16.57$^{\circ}$ &0.617 &0.373\\
    ManifoldMixup &0.42 &0.581 &13.76$^{\circ}$ &0.700 &0.304 &0.49 &0.460 &16.58$^{\circ}$ &0.588 &0.384\\
    Ours (PointMix) &0.41 &0.596 &13.77$^{\circ}$ &\textbf{0.711} &0.295 &0.48 &\textbf{0.508} &\textbf{14.34$^{\circ}$} &\textbf{0.611} &\textbf{0.360}\\
    CutMix &0.43 &0.578 &13.93$^{\circ}$ &0.693 &0.309 &0.49 &0.470 &15.66$^{\circ}$ &0.588 &0.380\\
    CutLossMix &0.40 &0.587 &\textbf{13.14$^{\circ}$} &0.676 &0.303 &\textbf{0.43} &0.473 &15.40$^{\circ}$ &0.559 &0.371\\
    Ours (PatchMix) &\textbf{0.40} &\textbf{0.615} &13.29$^{\circ}$ &0.690 &\textbf{0.292} &0.45 &0.486 &15.73$^{\circ}$ &0.576 &0.369\\
    \bottomrule
    \end{tabular}
    \vspace{-0.1cm}
\end{table*}

For video processing, we select 10 video frames per second evenly to predict the mouth keypoints of visible speakers. Then, the pixel coordinate of each person's mouth is used to generate two Gaussian-like vectors. Standard deviations are predefined with $\sigma_u^2=0.04$ and $\sigma_v^2=0.08$. The dimension of visual vectors is set to 64 to maintain consistency with audio features. For 10-second clips, the extracted visual features have dimensions of $100\times 64\times 12$. These visual features are then replicated five times along the temporal dimension to align with audio features, resulting in a fused audio-visual input with the shape of $500\times 64\times 19$.

ResNet-Conformer is employed as our network architecture. The parameters of ResNet are illustrated in Figure~\ref{fig3}. The kernel sizes for the three max-pooling operations in ResNet are $(1,4)$, $(1,4)$, and $(1,2)$, respectively. The embedding dimension input to the Conformer is set to $D=256$. The number of attention heads in Conformer is 8. Adam is used as the optimizer. The tri-stage learning rate scheduler \cite{park2019specaugment} is used with an upper limit of 0.001 for SELD-Teacher training. Given that SELD-Student is initialized with SELD-Teacher's parameters, we utilize a smaller learning rate of 0.0001 for the hierarchical cross-modal distillation process. 

According to our previous work \cite{jiang2024exploring}, for the task loss $\mathcal{L}^{\text{s}}$ and response-based KD loss $\mathcal{L}^{\text{rkd}}$, $\eta_1 = \beta_1 = 0.1$ are set as the loss weights for the SED subtask, and $\eta_2 = \beta_2 = 1.0$ are set as the loss weights for the SSL subtask, to balance the importance of the two subtasks during training. As shown in Fig.~\ref{fig:hcmd}, the feature-based KD method utilized $J=4$ stages of HCL loss. For $\mathcal{L}^{\text{fkd}}$, $M=4$ denotes the number of pooling layers, and the setting of $\alpha_m$ refers to \cite{chen2021distilling}. The overall training loss $\mathcal{L}$, as indicated by Equation (31), sets the weights of both distillation methods to 0.5. Additionally, the feature-based KD method employs a warm-up strategy of 20 epochs. 

The evaluation metrics adopted for the DCASE 2023 Challenge Task 3 include the location-dependent error rate (ER), location-dependent F-score (F), localization error (LE), and localization recall (LR). Both ER and F are calculated when the spatial error is within ${20^{\circ}}$. The overall evaluation metric for DCASE 2023 Task 3 is defined as follows,
\begin{equation}
{\text{Score}}=\frac{1}{4}[\text{ER}+(1-\text{F})+\frac{\text{LE}}{\pi}+(1-\text{LR})]
\end{equation}

The evaluation metrics for DCASE 2024 Task 3 include the location-dependent F1-score (F1), DOA error (DOAE), and relative distance error (RDE). The F1 is computed within a ${20^{\circ}}$ angular threshold and a 100\% relative distance threshold. The overall score for DCASE 2024 Task 3 is defined as follows,
\begin{equation}
{\text{Score}}=\frac{1}{3}[(1-\text{F1})+\frac{\text{DOAE}}{\pi}+\text{RDE}]
\end{equation}

\subsection{Results Based on Data Augmentation}
We systematically explore the effectiveness of various data augmentation methods for both audio and audio-visual SELD tasks in the DCASE 2023 Challenge. These methods differ in the mixing granularity and layer combinations, with results listed in Table~\ref{tab: table2}. The training set for the teacher model, which totals 190.4 hours, is a mixture of real data, synthetic data, and ACS-augmented samples. The student model was trained on a 30.4-hour dataset containing only real data along with augmentations generated by ACS and VPS. Both teacher and student models are trained using only the SELD task loss. Several key points emerge from the table.

First, classical data augmentation methods borrowed from image classification tasks, such as Mixup and CutMix, consistently improve performance in both audio-only and audio-visual SELD. Without using data augmentation, the baselines achieve overall scores of 0.330 and 0.456 for SELD-Teacher and SELD-Student, respectively, denoted as ``NoAug'' in the table. The Mixup technique yields relative improvements of 4.8\% and 17.3\% in overall metrics for the SELD-Teacher and SELD-Student models, respectively. CutMix achieves relative gains of 6.4\% and 16.7\% in the same models' overall metrics. Notably, the localization error LE metric shows limited improvement for the SELD-Teacher model, suggesting that the label-mixing operations used in Mixup and CutMix may be suboptimal for regression-based tasks like SSL.

Next, in the Mixup and CutMix methods, we propose replacing label mixing with loss mixing, resulting in modified approaches termed LossMix and CutLossMix. This adaptation proves more suitable for SSL tasks. Experimental results demonstrate consistent reductions in LE for both SELD-Teacher and SELD-Student models. Furthermore, the improved localization performance subsequently enhances SED metrics compared to conventional Mixup and CutMix approaches. With LossMix, the SELD-Teacher's overall metric improves from 0.314 to 0.305 compared to standard Mixup, showing gains across all four evaluation metrics. Using CutLossMix increases the SELD-Student's overall score from 0.380 to 0.371 versus traditional CutMix. These findings demonstrate that mixing losses rather than labels constitutes a more effective strategy for SELD tasks. 

\begin{table*}[t]
    \centering
    \caption{Performance comparison of the proposed PointMix and PatchMix for both 3D SELD and AV 3D SELD tasks evaluated on the development set of the DCASE 2024 Challenge Task 3.}
    \label{tab: table3} 
    \setlength{\tabcolsep}{9pt}
    \begin{tabular}{ccccc|cccc}
    \toprule
    \multirow{2}*{Method} & \multicolumn{4}{c}{\raisebox{0.3\height}{SELD-Teacher (Audio-Only)}} & \multicolumn{4}{c}{\raisebox{0.3\height}{SELD-Student (Audio-Visual)}} \\
    \cline{2-9}
    ~  
    & \raisebox{-0.3\height}{$\text{F1}\!\uparrow$} 
    & \raisebox{-0.3\height} {$\text{DOAE}\!\downarrow$} 
    & \raisebox{-0.3\height} {$\text{RDE}\!\downarrow$} 
    & \raisebox{-0.3\height} {$\text{Score}\!\downarrow$} 
    & \raisebox{-0.3\height}{$\text{F1}\!\uparrow$} 
    & \raisebox{-0.3\height} {$\text{DOAE}\!\downarrow$} 
    & \raisebox{-0.3\height} {$\text{RDE}\!\downarrow$} 
    & \raisebox{-0.3\height} {$\text{Score}\!\downarrow$}\\
    \midrule
    NoAug &0.452 &15.63$^{\circ}$ &0.271 &0.302 &0.325 &17.45$^{\circ}$ &0.271 &0.348\\
    Ours (PointMix) &0.509 &14.41$^{\circ}$ &\textbf{0.242} &\textbf{0.271} &0.429 & 15.13$^{\circ}$ &0.261 &0.305\\
    Ours (PatchMix) &\textbf{0.510} &\textbf{14.02}$^{\circ}$ &0.245 &\textbf{0.271} &\textbf{0.435} &\textbf{14.98}$^{\circ}$ &\textbf{0.257} &\textbf{0.302}\\
    \bottomrule
    \end{tabular}
    \vspace{-0.1cm}
\end{table*}

Finally, our proposed PointMix and PatchMix, which perform feature mixing at different hidden layers with varying granularities, achieve the most competitive results. While both methods demonstrate comparable performance, they significantly outperform the baseline system. Specifically, for the SELD-Teacher, PatchMix yields slightly better results than PointMix, improving the overall metric from 0.330 to 0.292 compared to the baseline without data augmentation. The SELD-Student with PointMix achieves a 21.1\% relative improvement in the overall metric, with a reduction in detection error ER from 0.58 to 0.48, an increase in detection F-score from 0.334 to 0.508, a decrease in localization error LE from 18.80° to 14.34°, and an enhancement in localization recall LR from 0.525 to 0.611 compared to the baseline. These results demonstrate that interpolations of hidden representations and loss mixing encourage the neural network to learn more discriminative features, thereby enhancing SELD performance. Additionally, ManifoldMixup surpasses traditional Mixup through its multi-layer mixing capability, extending beyond the input-layer restriction of the original method.

\begin{figure}
  \centering
  \begin{minipage}[b]{0.48\linewidth}
  \centering
  \includegraphics[width=0.995\linewidth]{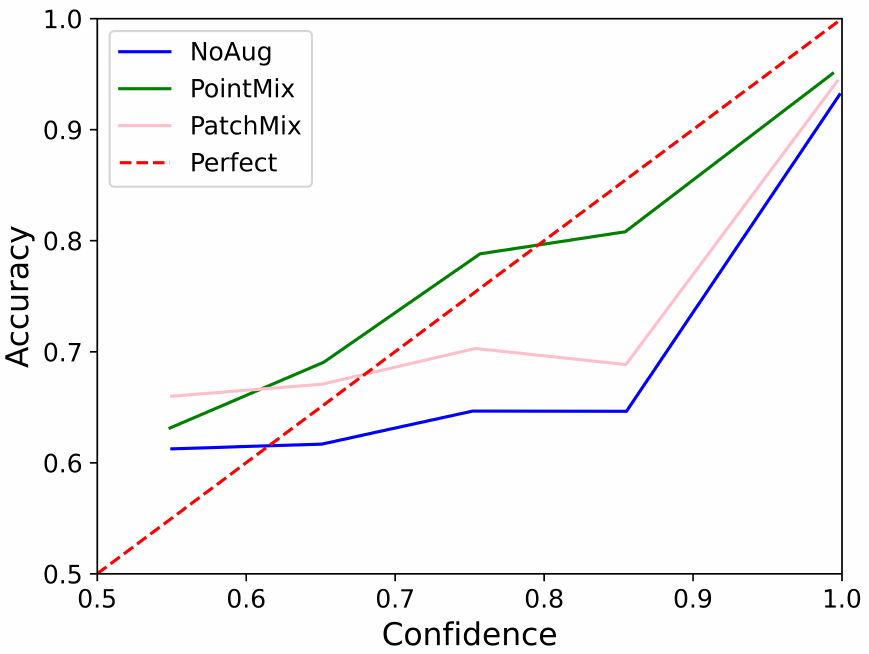}
  {\centerline{\scriptsize{(a) SELD-Teacher}}}
  \end{minipage}
  \label{fig:a-calib}
  \hfill
  \begin{minipage}[b]{0.48\linewidth}
  \centering
  \includegraphics[width=0.995\linewidth]{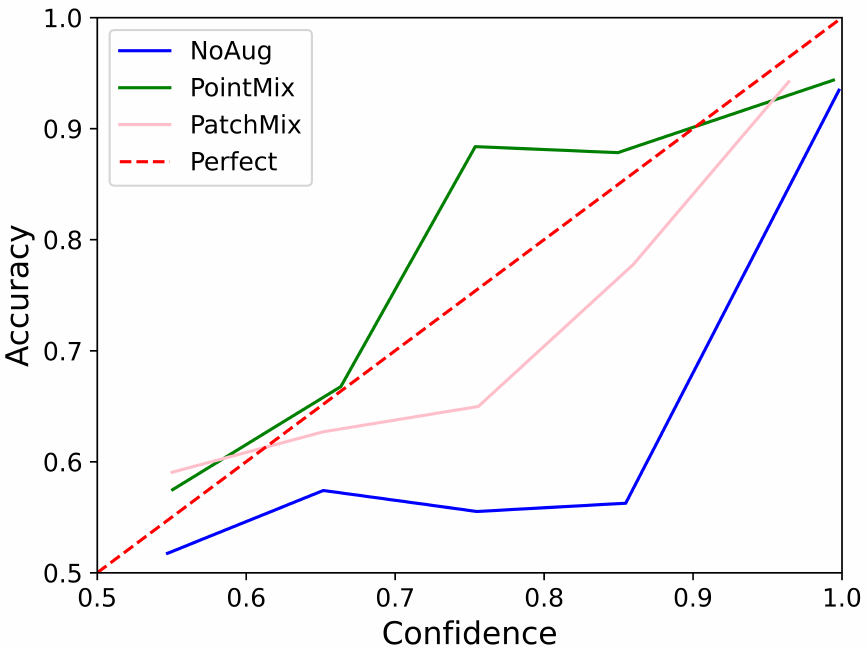}
  {\centerline{\scriptsize{(b) SELD-Student}}}
  \end{minipage}
  \label{fig:av-calib}

  \centering
  \caption{The calibration reliability plots for both (a) audio and (b) audio-visual models trained with and without data augmentation methods.}
  \label{fig:calibration}
\end{figure}

To further analyze the impact of data augmentation methods, we visualize the calibration reliability plots \cite{guo2017calibration}, as shown in Fig.~\ref{fig:calibration}. The reliability plots provide a qualitative assessment of the neural network's confidence calibration, depicting expected detection accuracy as a function of confidence. Fig.~\ref{fig:calibration} presents the reliability plots for both SELD-Teacher and SELD-Student models shown in Table~\ref{tab: table2} for the ``male speech'' sound class across different augmentation approaches. Perfect calibration is indicated by points adhering to the 45$^{\text {o}}$ reference line. Curves above the line suggest underconfidence, while those below indicate overconfidence. Given that the SED task involves binary classification, where a sound is considered active when the predicted probability exceeds a threshold of 0.5, the plot exclusively displays the confidence region above this threshold. The red dashed line indicates the ideal calibration. It is observed that the baseline systems, without the use of data augmentation methods, are prone to overconfidence. PointMix and PatchMix demonstrate consistent positive effects on the calibration of neural networks for sound event detection, indicating the reliability of models trained with these data augmentation methods. The visualization analysis corroborates the findings listed in Table~\ref{tab: table2}.

\begin{figure}
  \centering
  \includegraphics[width=0.6\linewidth]{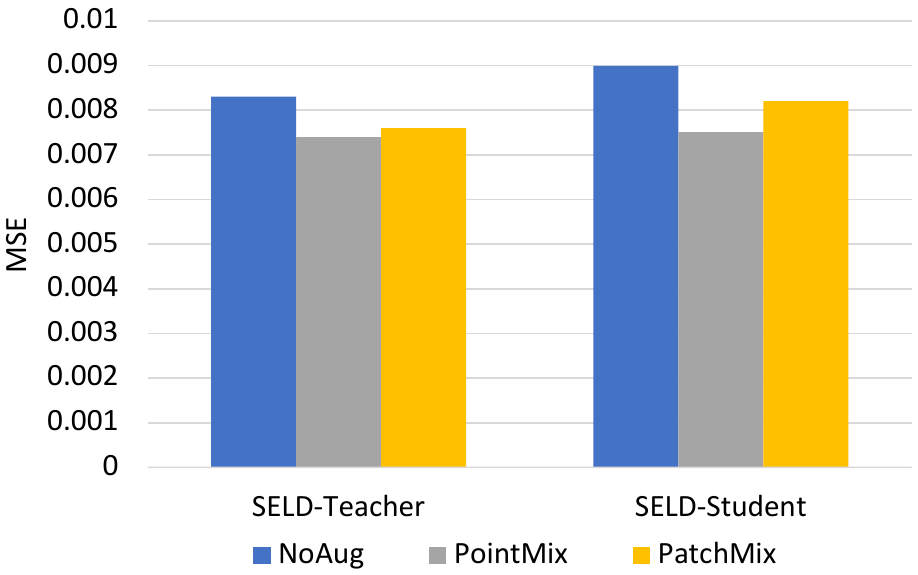}
  \caption{The MSE distribution across all sound event classes for both audio and audio-visual models trained with and without data augmentation methods.}
  \label{fig:mse_dist}
  \vspace{-0.1cm}
\end{figure}

Fig.~\ref{fig:mse_dist} illustrates the MSE distribution for the SSL task on the STARSS23 dataset for the SELD-Teacher and SELD-Student models trained with PointMix and PatchMix. From Fig.~\ref{fig:mse_dist}, it is evident that models trained without data augmentation consistently exhibit higher MSE in both audio-only and audio-visual systems. The proposed PointMix and PatchMix methods address this limitation by generating additional training samples through linear interpolation of hidden representations and losses, which guide the optimization of SELD tasks. This dual mechanism of hidden-space augmentation and loss-based regularization greatly enhances model robustness, resulting in lower MSE compared to the baseline system, which aligns with the localization metrics reported in Table~\ref{tab: table2}.

As shown in Table~\ref{tab: table2}, PointMix and PatchMix yield superior performance among various data augmentations. Consequently, we conducted experiments using these two methods for both 3D SELD and AV 3D SELD in the DCASE 2024 Challenge Task 3, with results presented in Table~\ref{tab: table3}. The DCASE 2024 Challenge introduced a novel requirement for SELD systems to estimate both sound source directions and distances, significantly increasing the task's difficulty. While the baseline of the DCASE 2023 Challenge Task 3 achieves an F score of 0.540 and an LE value of 14.34°, extending to 3D SELD method reduces the F1 to 0.452 and increases the DOAE value to 15.63°, highlighting the challenges introduced by distance estimation. Our proposed PointMix and PatchMix methods demonstrate consistent performance gains in audio (teacher model) and audio-visual (student model) 3D SELD tasks, with improvements across all three evaluation metrics.

\subsection{Results Based on Hierarchical Cross-modal Distillation}

In this subsection, we conduct experiments on hierarchical cross-modal distillation. Firstly, we present the performance comparison among different data setups for teacher and student models in Table~\ref{tab: table4}. Systems T0 to T2 and S1 to S2 denote teacher and student models trained with varying data sizes. System T0 denotes the audio model trained with the STARSS23 dataset, yielding slightly worse performance compared to the audio-visual model S1 under matched training conditions. The results for T0 to T2, S1 to S2, and the RKD methods are reported from our previous work \cite{jiang2024exploring}. 

\begin{table}
    \centering
    \caption{Performance comparison among different data setups for the teacher models T1 to T2 and student models S1 to S2 presented in Tabel~\ref{tab: table5}. The results are reported from \cite{jiang2024exploring}.}
    \label{tab: table4}
    \setlength{\tabcolsep}{4pt}
    \begin{tabular}{cccccccc}
    \toprule
    \raisebox{-0.3\height} {System}
    & \raisebox{-0.3\height} {Data set}
    & \raisebox{-0.3\height} {Data size}
    & \raisebox{-0.3\height} {$\text{ER}\!\downarrow$} 
    & \raisebox{-0.3\height}{$\text{F}\!\uparrow$} 
    & \raisebox{-0.3\height} {$\text{LE}\!\downarrow$} 
    & \raisebox{-0.3\height} {$\text{LR}\!\uparrow$} 
    & \raisebox{-0.3\height} {$\text{Score}\!\downarrow$} \\
    \midrule
    Teacher &\multicolumn{7}{c}{} \\
    T0 &Real &3.8h &0.75 &0.170 &39.82$^{\circ}$ &0.380 &0.605 \\
    T1 &Real+Synth &23.8h &0.56 &0.420 &18.65$^{\circ}$ &0.670 &0.393 \\
    T2 &+ACS &190.4h &0.45 &0.550 &14.28$^{\circ}$ &0.660 &0.330 \\
    \midrule
    Student &\multicolumn{7}{c}{} \\
    S1 &Real &3.8h &0.76 &0.180 &34.13$^{\circ}$ &0.420 &0.587 \\
    S2 &+ACS+VPS &30.4h &0.57 &0.360 &18.82$^{\circ}$ &0.510 &0.451\\
    \bottomrule
    \end{tabular}
    \vspace{-0.1cm}
\end{table}

\begin{table}
    \centering
    \caption{Performance comparison of the RKD and HCMD methods applied to the AV SELD systems trained with different data sizes in the DCASE 2023 Challenge Task 3.}
    \label{tab: table5} 
    \setlength{\tabcolsep}{7pt}
    \begin{tabular}{ccccccc}
    \toprule
    Method & System & \raisebox{-0.3\height} {$\text{ER}\!\downarrow$} 
    & \raisebox{-0.3\height}{$\text{F}\!\uparrow$} 
    & \raisebox{-0.3\height} {$\text{LE}\!\downarrow$} 
    & \raisebox{-0.3\height} {$\text{LR}\!\uparrow$} 
    & \raisebox{-0.3\height} {$\text{Score}\!\downarrow$} \\
    \midrule
    \multirow{3}{*}{RKD} &
    T1-S1 &0.64 &0.280 &33.93$^{\circ}$ &0.574 &0.494\\
    ~ &T2-S1 &0.51 &0.394 &28.37$^{\circ}$ &0.581 &0.423\\
    ~ &T2-S2 &0.44 &0.516 &12.80$^{\circ}$ &0.597 &0.350\\
    \midrule
    \multirow{3}{*}{HCMD} &
    T1-S1 &0.56 &0.359 &21.56$^{\circ}$ &0.544 &0.444\\
    ~ &T2-S1 &0.49 &0.477 &17.63$^{\circ}$ &0.596 &0.379\\
    ~ &T2-S2 &0.44 &0.532 &13.27$^{\circ}$ &0.616 &0.341\\
    \bottomrule
    \end{tabular}
    \vspace{-0.1cm}
\end{table}

\begin{table*}[t]
    \centering
    \caption{Performance comparison of teacher and student models with and without the HCMD method when using PointMix and PatchMix.}
    \label{tab: table6}
    \setlength{\tabcolsep}{9pt}
    \begin{tabular}{ccccccc|cccc}
    \toprule
    \multirow{2}*{Method} &\multirow{2}*{System} &\multicolumn{5}{c}{\raisebox{0.3\height}{DCASE 2023 Challenge Task 3}} & \multicolumn{4}{c}{\raisebox{0.3\height}{DCASE 2024 Challenge Task 3}} \\
    \cline{3-11}
    ~  &~ & \raisebox{-0.3\height} {$\text{ER}\!\downarrow$} 
    & \raisebox{-0.3\height}{$\text{F}\!\uparrow$} 
    & \raisebox{-0.3\height} {$\text{LE}\!\downarrow$} 
    & \raisebox{-0.3\height} {$\text{LR}\!\uparrow$} 
    & \raisebox{-0.3\height} {$\text{Score}\!\downarrow$} 
    & \raisebox{-0.3\height}{$\text{F1}\!\uparrow$} 
    & \raisebox{-0.3\height} {$\text{DOAE}\!\downarrow$} 
    & \raisebox{-0.3\height} {$\text{RDE}\!\downarrow$} 
    & \raisebox{-0.3\height} {$\text{Score}\!\downarrow$}\\
    \midrule
    \multirow{3}{*}{PointMix}
    &Teacher &0.41 &0.596 &13.77$^{\circ}$ &0.711 &0.295 &0.509 &14.41$^{\circ}$ &\textbf{0.242} &0.271\\
    ~ &Student &0.48 &0.508 &14.34$^{\circ}$ &0.611 &0.360 &0.429 & 15.13$^{\circ}$ &0.261 &0.305 \\
    ~ &+HCMD  &\textbf{0.41} &\textbf{0.616} &\textbf{13.50$^{\circ}$} &\textbf{0.733} &\textbf{0.284} &\textbf{0.551} &\textbf{13.83$^{\circ}$} &0.255 & \textbf{0.260}\\
    \midrule
    \multirow{3}{*}{PatchMix} 
    &Teacher &0.40 &0.615 &\textbf{13.29$^{\circ}$} &\textbf{0.690} &\textbf{0.292} &\textbf{0.510} &14.02$^{\circ}$ &\textbf{0.245} &\textbf{0.271}\\
    ~ &Student &0.45 &0.486 &15.73$^{\circ}$ &0.576 &0.369 &0.435 &14.98$^{\circ}$ &0.257 &0.302\\
    ~ &+HCMD  &\textbf{0.38} &\textbf{0.616} &13.38$^{\circ}$ &0.660 &0.295 &0.506 &\textbf{13.59$^{\circ}$} &0.252 &0.274\\
    \midrule
    - &Student &0.58 &0.334 &18.80$^{\circ}$ &0.525 &0.456 &0.325 &17.45$^{\circ}$ &0.271 &0.348\\
    \bottomrule
    \end{tabular}
    \vspace{-0.1cm}
\end{table*}

We then analyze the effects of training data size on two knowledge distillation methods and present the results in Table~\ref{tab: table5}. ``RKD'' denotes the application of the response-based knowledge distillation method, while ``HCMD'' refers to the proposed hierarchical cross-modal distillation method, which utilizes both the response-based and feature-based knowledge distillation techniques. The student model trained with a small dataset is guided by the teacher model trained on a larger dataset. For example, the notation ``T1-S1'' represents the optimized system where the S1 student model is guided by the T1 teacher model. Under each guidance configuration, the student model employing the RKD method consistently outperforms its non-guided counterpart. 

Building upon RKD, extracting knowledge from the teacher's hidden layer features further enhances the student model's performance, especially for the `T1-S1'' and `T2-S1'' systems. For instance, as shown in Table~\ref{tab: table4}, the S1 student, trained on real data, achieved an overall score of 0.587. Employing RKD improved the overall score to 0.494, with consistent increases across all four metrics. The proposed HCMD method further enhanced the overall metric to 0.444. This demonstrates that intermediate-layer knowledge from the teacher model is highly effective in boosting the performance of the student model, especially when the latter has access to only a very limited amount of training data. While the teacher model does not process visual inputs, it learns highly discriminative acoustic representations from its training on a large audio dataset. The student model, initialized with the teacher's parameters, leverages this strong starting point as it fine-tunes on fused audio-visual features while simultaneously being guided by the teacher's audio expertise. Notably, when using the T2 teacher to guide the S2 student, our proposed HCMD method achieves SELD performance nearly comparable to that of the T2 teacher model, with even superior results in detection error rate and localization error metrics.

\begin{figure}
  \centering
  \begin{minipage}[b]{0.368\linewidth}
  \centering
  \includegraphics[width=0.998\linewidth]{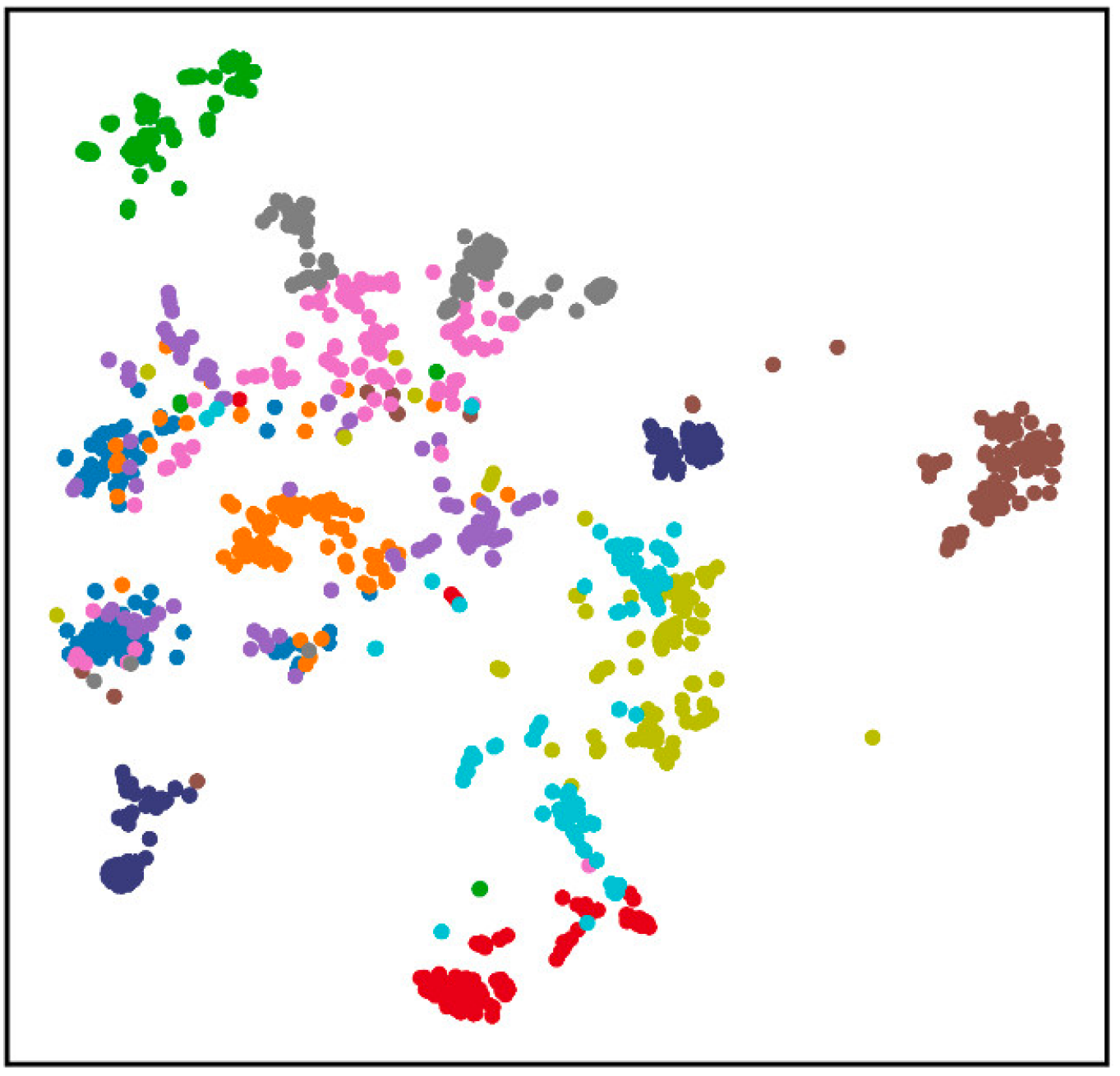}
  \end{minipage}
  \label{fig:fig10}
  \hfill
  \begin{minipage}[b]{0.618\linewidth}
  \centering
  \includegraphics[width=0.998\linewidth]{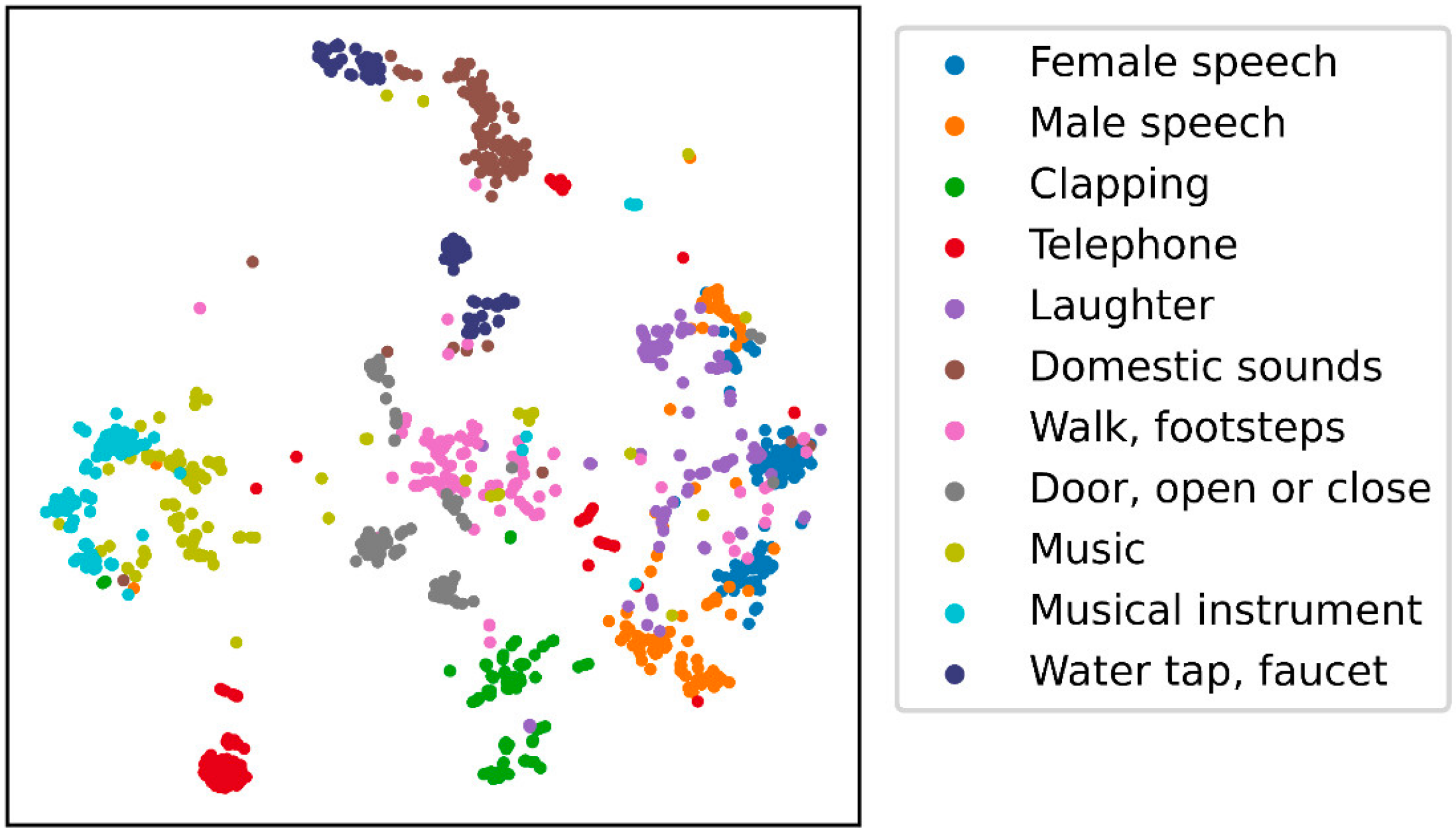}
  \end{minipage}
  \label{fig:fig11}

  \centering
  \vspace{-0.3cm}
  \caption{T-SNE visualization of embeddings from two students trained with only RKD (left) and with a combination of RKD and FKD (right), respectively.}
  \label{fig:t-sne}
  \vspace{-0.1cm}
\end{figure}

To understand the features we learned from the two KD methods, we visualize the t-SNE of hidden layer embeddings from the Conformer for the two ``T1-S1'' student models presented in Table~\ref{tab: table5}. For each sound class, we randomly selected 100 audio frames that contain only that specific class of sound, excluding any frames with concurrent sounds. The embeddings for bell and knock were omitted as they consistently co-occur with other sounds. As illustrated in Fig.~\ref{fig:t-sne}, for the RKD method, the class centers of similar sounds like telephone, music, and musical instrument are scattered and tend to cluster together. However, after applying FKD, the centers for these three classes become more compact and clearly distinguished. This indicates that with guidance from the teacher model's intermediate representations, the student model is able to learn discriminative features. Laughter is another class that is difficult to distinguish. The laughter embeddings produced by the RKD method are scattered among those for male speech, female speech, and footsteps. Laughter embeddings from a combination of the RKD and FKD methods cluster more compactly, achieving a clearer separation from the footsteps class overall. Nevertheless, the embeddings for laughter remain in close to both male and female speech, which is attributable to their frequent co-occurrence in the audio data.

Subsequently, multi-level data augmentation is combined with the HCMD method to further improve the SELD performance. The results of the DCASE 2023 and 2024 Challenge Task 3 are shown in Table~\ref{tab: table6}. We conducted statistical significance tests on the overall scores of the student models before and after applying HCMD, with all p-values less than 0.05. Regardless of whether PointMix or PatchMix is employed, the proposed HCMD framework demonstrates significant superiority over the original student model, achieving better performance than the teacher model across the majority of evaluation metrics. For instance, with PointMix, the teacher model achieves an overall score of 0.295, while the student model scores 0.360. By utilizing the proposed HCMD method, the student model outperforms the teacher model, with the score improving from 0.295 to 0.284. PointMix yields sightly better performance than PatchMix. The experimental results demonstrate the effectiveness of the unified framework that combines multi-level data augmentation and hierarchical cross-modal distillation methods.

\begin{table}[t]
    \centering
    \caption{Performance comparison of PointMix for different sets on eligible layers and mixing coefficients for AV SELD on the development set of the DCASE 2023 Challenge Task 3.}
    \label{tab: table7}
    \begin{tabular}{cccccccc}
    \toprule
    Layer
    & \raisebox{-0.3\height} {$D$}
    & \raisebox{-0.3\height} {$\alpha$}
    & \raisebox{-0.3\height} {$\text{ER}\!\downarrow$} 
    & \raisebox{-0.3\height}{$\text{F}\!\uparrow$} 
    & \raisebox{-0.3\height} {$\text{LE}\!\downarrow$} 
    & \raisebox{-0.3\height} {$\text{LR}\!\uparrow$} 
    & \raisebox{-0.3\height} {$\text{Score}\!\downarrow$} \\
    \midrule
    \multirow{3}{*}{Conv} &\multirow{3}{*}{19} &0.5 &0.52 &0.394 &19.95$^{\circ}$ &0.587 &0.411 \\
    ~ &~ &1.0 &0.52 &0.403 &15.66$^{\circ}$ &0.562 &0.409\\
    ~  &~ &2.0 &0.52 &0.444 &17.33$^{\circ}$ &0.601 &0.393\\
    \midrule
    \multirow{3}{*}{BasicBlock} &\multirow{3}{*}{9} &0.5 &0.52 &0.434 &17.44$^{\circ}$ &0.589 &0.399 \\
    ~ &~ &1.0 &0.51 &0.451 &16.32$^{\circ}$ &0.596 &0.388\\
    ~  &~ &2.0 &0.48 &0.450 &14.92$^{\circ}$ &0.571 &0.384\\
    \midrule
    \multirow{3}{*}{ResBlock} &\multirow{3}{*}{5} &0.5 &0.48 &0.493 &14.96$^{\circ}$ &0.593 &0.369 \\
    ~ &~ &1.0 &\textbf{0.48} &\textbf{0.508} &\textbf{14.34$^{\circ}$} &\textbf{0.611} &\textbf{0.360}\\
    ~  &~ &2.0 &0.49 &0.466 &15.98$^{\circ}$ &0.622 &0.373\\
    \bottomrule
    \end{tabular}
    \vspace{-0.1cm}
\end{table}

\subsection{Ablation Study on Mixing Layers}
We conduct a series of ablation experiments by using different sets on eligible layers and varying mixing coefficients to perform interpolations of hidden representations. Table~\ref{tab: table7} presents the experimental results of PointMix applied to the student model. Variable $D$ denotes the number of eligible layers for feature mixing, and $\alpha$ is the parameter of Beta distribution. We explore three sets of eligible layers within the ResNet architecture, each marked with different colors in Fig.~\ref{fig2}. The set of eligible layers includes the input layer. When implementing mixing after individual convolutional layers, there are 19 eligible layers for combination. This granularity decreases when operating at higher structural levels - mixing performed after BasicBlocks offers 9 candidate layers, while mixing at the ResBlock level further reduces the number of eligible layers to just 5. Compared to the student model, which achieves an overall score of 0.456 without using augmentation as shown in Table~\ref{tab: table2}, PointMix demonstrates superior SELD performance. PointMix yields the best results when performing interpolation after ResBlocks. This hierarchical reduction of eligible layers reflects the progressive enhancement of feature mixing through deeper network architecture stages. Therefore, in this study, both PointMix and PatchMix perform interpolation of hidden representations by randomly selecting candidates from a set comprising the input layer and four residual blocks. Additionally, The selection of the mixing coefficient, determined by the Beta distribution, influences the feature mixing ratio in the data augmentation method, thereby affecting the final AV SELD performance. In this study, all models are trained by selecting a proper hyper-parameter $\alpha$.

\subsection{Comparison with State-of-the-Art Methods}
\begin{table}[t]
\centering
\caption{Performance comparison with SOTA methods on the development set of the DCASE 2023 Challenge Task 3.}
\label{tab:table8}
\setlength{\tabcolsep}{9pt}
\begin{tabular}{lccccc}
\toprule
System & \raisebox{-0.3\height} {$\text{ER}\!\downarrow$} 
    & \raisebox{-0.3\height}{$\text{F}\!\uparrow$} 
    & \raisebox{-0.3\height} {$\text{LE}\!\downarrow$} 
    & \raisebox{-0.3\height} {$\text{LR}\!\uparrow$} 
    & \raisebox{-0.3\height} {$\text{Score}\!\downarrow$} \\
\midrule
Baseline \cite{shimada2023starss23} &1.07 & 0.143 & 48.00$^{\circ}$ & 0.355 & 0.710 \\
Berghi \cite{berghi2024fusion} &0.51 & 0.495 & 15.80$^{\circ}$ & 0.602 &0.375 \\
Kang \cite{Kang_KT_task3_report} &0.45 & 0.494 & 16.00$^{\circ}$ & 0.637 &0.352 \\
Jiang \cite{jiang2024exploring} &0.41 &0.590 &14.10$^{\circ}$ &0.730 &0.292 \\
HDA-SELD &\textbf{0.41} & \textbf{0.616} & \textbf{13.50$^{\circ}$} & \textbf{0.733} & \textbf{0.284} \\
\bottomrule
\end{tabular}
\vspace{-0.1cm}
\end{table}

We compare our proposed method with state-of-the-art (SOTA) methods on two AV SELD tasks of the DCASE 2023 and 2024 Challenges. The results for the AV SELD task of the DCASE 2023 Challenge are presented in Table~\ref{tab:table8}. The baseline system was trained solely with the STARSS23 dataset, which contains about 3.8 hours of real data, as shown in Table~\ref{tab: table4}. The baseline system utilized visual embeddings similar to ours and adopted a lightweight network as the backbone. Kang's system was trained with a larger dataset than the remaining systems, employing video mosaic augmentation. YOLOv7 \cite{wang2023yolov7} was used as the visual encoder in Kang's system. To ensure a fair comparison, the systems by Berghi, Jiang, and our proposed HDA-SELD were all trained on a dataset eight times larger than the original STARSS23 data. This eightfold increase was achieved by consistently applying similar video augmentation techniques in addition to the ACS transformation, as shown in Table~\ref{tab: table4}. It is observed that our proposed HDA-SELD outperforms all other systems in the overall metric score, with a 0.026 improvement in F and a 0.6 decrease in LE compared to the previous SOTA system.

\begin{table}[t]
\centering
\caption{Performance comparison with SOTA methods on the development set of the DCASE 2024 Challenge Task 3.}
\label{tab:table9}
\setlength{\tabcolsep}{10pt}
\begin{tabular}{lcccc}
\toprule
System & \raisebox{-0.3\height}{$\text{F1}\!\uparrow$} 
    & \raisebox{-0.3\height} {$\text{DOAE}\!\downarrow$} 
    & \raisebox{-0.3\height} {$\text{RDE}\!\downarrow$} 
    & \raisebox{-0.3\height} {$\text{Score}\!\downarrow$}\\
\midrule
Baseline \cite{krause2024sound} & 0.113 & 38.40$^{\circ}$ & 0.360 & 0.487 \\
Li \cite{Li_SHU_task3b_report} & 0.392 & 18.70$^{\circ}$ & 0.310 & 0.341 \\
Berghi \cite{Berghi_SURREY_task3b_report}& 0.408 & 17.70$^{\circ}$ & 0.305 & 0.332 \\
Hong \cite{hong2025mvanet}& 0.495 & 13.90$^{\circ}$ & \textbf{0.245} & 0.276 \\
HDA-SELD & \textbf{0.551} & \textbf{13.83$^{\circ}$} & 0.255 & \textbf{0.260} \\
\bottomrule
\end{tabular}
\vspace{0.1cm}
\end{table}

The results on the AV 3D SELD task of the DCASE 2024 Challenge are shown in Table~\ref{tab:table9}. The baseline was also trained with the original STARSS23 dataset; whereas, all remaining systems were trained with a larger data set containing about 30.8 hours of training data. Unlike all comparative systems that utilized ResNet-50 \cite{he2016deep} for visual encoding, our approach used Gaussian-like vectors as the visual embedding. In Li's system \cite{Li_SHU_task3b_report}, audio and visual features were fed into a Transformer decoder for feature fusion utilizing cross attention. Berghi's method \cite{Berghi_SURREY_task3b_report} employed a Conformer module for processing the concatenated multi-modal embeddings. Hong et al. \cite{hong2025mvanet} proposed a multi-stage video attention network to address the AV 3D SELD task. Our proposed HDA-SELD achieves the best overall metric score of 0.260. 
Specifically, our proposed method increases the F1 score from 0.495 to 0.551, an 11.3\% improvement compared to Hong's model, which highlights the effectiveness of our approach in addressing the complex AV 3D SELD task.

\section{Discussion}

Mouth region related Gaussian vectors are used as visual features in this study. A key limitation of this approach is its inability to represent non-vocal sound sources, such as music, running water, and footsteps. To address this, our future work will explore using a multi-modal large language model (MLLM), Grounding-GPT \cite{li2024groundinggpt}, to extract fine-grained visual features for all target sound categories. 

Although the evaluation was validated on small-scale datasets, the proposed knowledge distillation and data augmentation methods would be applied to other multi-modal tasks and datasets, such as audio-visual speech recognition \cite{zmolikova2024chime} and visual sound localization \cite{senocak2018learning}. For instance, the proposed knowledge distillation approach can be leveraged to transfer expertise from a large-scale speech-trained model to an audio-visual student model. Additionally, the multi-level data augmentation strategy is also applicable to visual sound localization tasks that are addressed via coordinate regression.

\section{Conclusion}
This study focuses on hierarchical cross-modal distillation and multi-level data augmentation for AV SELD. The proposed hierarchical cross-modal distillation aims to transfer rich information from both the output responses and hidden-layer features of the teacher model to guide the training of the student model. Multi-level data augmentation methods involve randomly selecting representations from multiple network layers for linear interpolation, and mixing loss functions instead of labels. This approach is particularly suitable for SELD tasks, which include regression objectives, and effectively enhances the diversity of training samples. Tested on the DCASE 2023 and 2024 Challenge Task 3 datasets, our proposed HDA-SELD, which combines hierarchical cross-modal distillation and multi-level data augmentation, has achieved significant performance improvements, outperforming our first-place systems in both challenges.

\small
\bibliographystyle{IEEEtran}
\bibliography{IEEEabrv,ref}

\end{document}